\documentclass[hyper,a4paper]{SIMO} 
\usepackage{graphicx,graphics}
\usepackage{cite,citesort}
\usepackage{rotating,rotate}
\usepackage{epsfig}
\usepackage{exscale}
\usepackage{amsmath,amssymb}
\usepackage{cite}
\usepackage{multirow}
\usepackage{latexsym}
\usepackage[english]{babel}
\newcommand{\bc}{\begin{center}}
\newcommand{\ec}{\end{center}}
\newcommand{\bfi}{\begin{figure}}
\newcommand{\efi}{\end{figure}}
\newcommand{\bm}{\begin{minipage}}
\renewcommand{\em}{\end{minipage}}
\newcommand{\be}{\begin{equation}}
\newcommand{\en}{\end{equation}}
\newcommand{\bea}{\begin{eqnarray}}
\newcommand{\eea}{\end{eqnarray}}
\newcommand{\bi}{\begin{itemize}}
\newcommand{\ei}{\end{itemize}}
\newcommand{\no}{\nonumber}

\newcommand{\gs}{g^{_S}_{_{12}}}

\newcommand{\ov}[1]{\overline{#1}}
\newcommand{\bbr}{{\it B{\footnotesize A}B{\footnotesize AR}}}
\def\b         {\ensuremath{\mathcal{B}}}
\def\bb        {\ensuremath{\mathcal{B}\overline{\mathcal{B}}}}
\def\eepp        {\ensuremath{e^+ e^- \!\rightarrow p\overline{p}}}
\def\eennb        {\ensuremath{e^+ e^- \!\rightarrow n\overline{n}}}
\def\eett        {\ensuremath{e^+ e^- \!\rightarrow \tau^+\tau^-\,}}

\def\eebb        {\ensuremath{e^+ e^- \!\rightarrow \mathcal{B}\mathcal{\overline{B}}}}
\def\eell        {\ensuremath{e^+ e^- \!\rightarrow \Lambda\overline{\Lambda}}}
\def\eelclc      {\ensuremath{e^+ e^- \!\rightarrow \Lambda_c\overline{\Lambda_c}}}
\def\eenns      {\ensuremath{e^+ e^- \!\rightarrow p\ov{N}(1440)}}
\def\lc        {\ensuremath{\Lambda_c}}

\def\sisi        {\ensuremath{\Sigma^0\overline{\Sigma^0}}}
\def\ls        {\ensuremath{\Lambda\overline{\Sigma^0}}}

\def\pp        {\ensuremath{p\overline{p}}}
\def\ll        {\ensuremath{\Lambda\overline{\Lambda}}}
\def\lclc        {\ensuremath{\Lambda_c\overline{\Lambda_c}}}
\def\ee        {\ensuremath{e^+e^-}}
\def\nb        {\ensuremath{{\rm nb}}}

\def\gev       {\ensuremath{{\rm GeV}}}
\def\mev       {\ensuremath{{\rm MeV}}}
\def\ge        {\ensuremath{G_E}}
\def\gm        {\ensuremath{G_M}}
\def\gs        {\ensuremath{G_S}}
\def\gd        {\ensuremath{G_D}}
\def\geb        {\ensuremath{G_E^\mathcal{B}}}
\def\gmb        {\ensuremath{G_M^\mathcal{B}}}
\def\gsb        {\ensuremath{G_S^\mathcal{B}}}
\def\gdb        {\ensuremath{G_D^\mathcal{B}}}
\title{%
No Sommerfeld resummation factor in \eepp\ ?
}
\author{Rinaldo Baldini Ferroli$^{\rm a,b}$, Simone Pacetti$^{\rm a,b}$, 
Adriano Zallo$^{\rm b}$ \\
 $^{\rm a}$Museo Storico della Fisica e Centro Studi e Ricerche ``E. Fermi'', Rome, Italy \\
 $^{\rm b}$INFN, Laboratori Nazionali di Frascati, Frascati, Italy  \\ 
\hspace{-6mm}\begin{tabular}{ll}
E-mail: &\email{baldini@centrofermi.it}\\
 &\email{simone.pacetti@lnf.infn.it}\\ 
&\email{adriano.zallo@lnf.infn.it}\\
\end{tabular}%
}
\abstract{%
The Sommerfeld rescattering formula is compared to the \eepp\ \bbr\ data
at threshold and above. While there is the expected Coulomb enhancement at threshold,
two unexpected outcomes have been found: $|G^p (4M_{p}^2)|= 1$, like for a pointlike fermion,
 and moreover data show that the resummation factor in 
the Sommerfeld formula is not needed. 
Other $\ee\to$ baryon-antibaryon cross sections show a similar behavior near threshold.
}
\begin{document}
Many recent papers, mostly concerning evidence of Dark Matter~\cite{darkm}, 
are related to the so called Sommerfeld rescattering formula~\cite{somm} [eq.~(\ref{eq:C})].
In this letter the unexpected lack of the resummation term in the 
Sommerfeld rescattering formula in the present \eepp\ cross section data, as well as 
other unexpected features of  \eebb\ (\b\ stands for baryon), are emphasized, 
namely:
\begin{itemize}
\item in \eepp\ the cross section is not vanishing at threshold, as 
already pointed out \cite{noi1,noi2}, fully dominated by the Coulomb 
final state enhancement and $|G^p(4 M_p^2) |= 1$,
where $G^p$ is the common value of electric and magnetic proton form factors
at the production threshold;
\item in \eepp\ the cross section above threshold is consistent with no 
Sommerfeld resummation factor, as it will be shown in detail in the following;
\item other charged baryon pair  cross sections data, \eelclc\ and \eenns\
+ c.c., show similar features, even if within large errors, as well as the 
present puzzling data on neutral baryon pair cross sections.
\end{itemize}
These cross sections have been measured by means of initial state radiation. 
This technique has several advantages in the case of pair production; 
in particular at threshold in the pair center of mass (c.m.) energy:
\begin{itemize}
\item the efficiency is quite high;
\item a very good invariant mass resolution is achieved, 
$\Delta W_{p\overline{p}} \sim 1$  \mev, comparable to what is obtained in a 
symmetric storage ring. 
\end{itemize}
In Born approximation the cross section for the process \eebb\ in the case of 
charged baryon pairs is assumed to be:
\bea
\sigma_{\bb}(W_{\bb}^2)=
\frac{4\pi\alpha^2\beta}{3W_{\bb}^2}\,\mathcal{C}
\left[
\left|\gmb(W_{\bb}^2)\right|^2+\frac{2M_\b^2}{W_{\bb}^2}\left|\geb(W_{\bb}^2)\right|^2
\right]\,,
\label{eq:sigma}
\eea
where $W_{\bb}$ is the \bb\ invariant mass, 
$\beta$ is the velocity of the outgoing baryon, $\mathcal{C}$ is the Coulomb
factor
\be
\mathcal{C}=\frac{\pi\alpha/\beta}{1-\exp(-\pi\alpha/\beta)}\,,
\label{eq:C}
\en
that takes into account the electromagnetic \bb\ final state interaction, 
$G_M^\b$ and $G_E^\b$ are the 
magnetic and electric Sachs form factors (FF).
\newline
Because of parity conservation S and D wave only are allowed. At threshold it is 
assumed that, according to the analyticity of the Dirac and Pauli FF's
as well as according to the S-wave dominance, there is only one FF: 
$G_E^\b(4 M^2_\b) = G_M^\b(4 M^2_\b) \equiv G^\b(4 M^2_\b)$.
The relationship between \geb, \gmb\ and \gsb, \gdb, the partial-wave FF's, are:
\bea
\gsb=\frac{2\gmb\sqrt{W_{\bb}^2/4M_\b^2}+\geb}{3}\,,\hspace{5mm}
\gdb=\frac{\gmb\sqrt{W_{\bb}^2/4M_\b^2}-\geb}{3}\,.
\label{eq:gsgd}
\eea
and $\gdb(4 M^2_\b) = 0$. For pointlike fermions it is
$G_E(Q^2) = G_M(Q^2) \equiv 1$. 
\newline
The Coulomb factor, $\mathcal{C}$, is usually introduced as an enhancement factor 
$\mathcal{E}$ times a resummation term $\mathcal{R}$, i.e. the so called 
Sommerfeld-Schwinger-Sakharov rescattering formula~\cite{coulomb,somm}: 
$\mathcal{C}=\mathcal{E} \times \mathcal{R}$.
It has a very weak dependence on the fermion pair total spin and corresponds to the squared 
value of the Coulomb scattering wave function at the origin, assumed as a good  
approximation of the Coulomb final state interaction. 
In such an approximation the factor $\mathcal{C}$ should affect the S wave 
only, because the D wave vanishes at the origin. For the same reason 
a Coulomb enhancement is not expected when pseudoscalar meson pairs are produced
via \ee\ annihilation; in fact these processes occur only in P wave.
The cross section formula should be more properly written in terms of S and 
D wave FF's:
\bea
\sigma_{\bb}(W_{\bb}^2)=
2\pi\alpha^2\beta\,\frac{4M_\b^2}{W_{\bb}^4}
\Big[
\mathcal{C}\,|\gsb(W_{\bb}^2)|^2+2|\gdb(W_{\bb}^2)|^2
\Big]\,.
\label{eq:sigma2}
\eea
 The enhancement factor is
\bea
 \mathcal{E} = \frac{\pi\alpha}{\beta}\,,\no
\eea
so that the phase space factor $\beta$ is cancelled and the cross 
section is expected to be finite and not vanishing even exactly at threshold.
As the resummation factor is
\bea
\mathcal{R} = \frac{1}{1-\exp\left(-\pi\alpha/\beta\right)}\,,
\label{eq:resum}
\eea
it follows that few MeV above the threshold it is $\mathcal{C} \sim 1$, the phase space factor is 
restored and Coulomb effects can be neglected.
\bfi[h!]\vspace{-2mm}
\bc
\bm{85mm}\bc
\epsfig{file=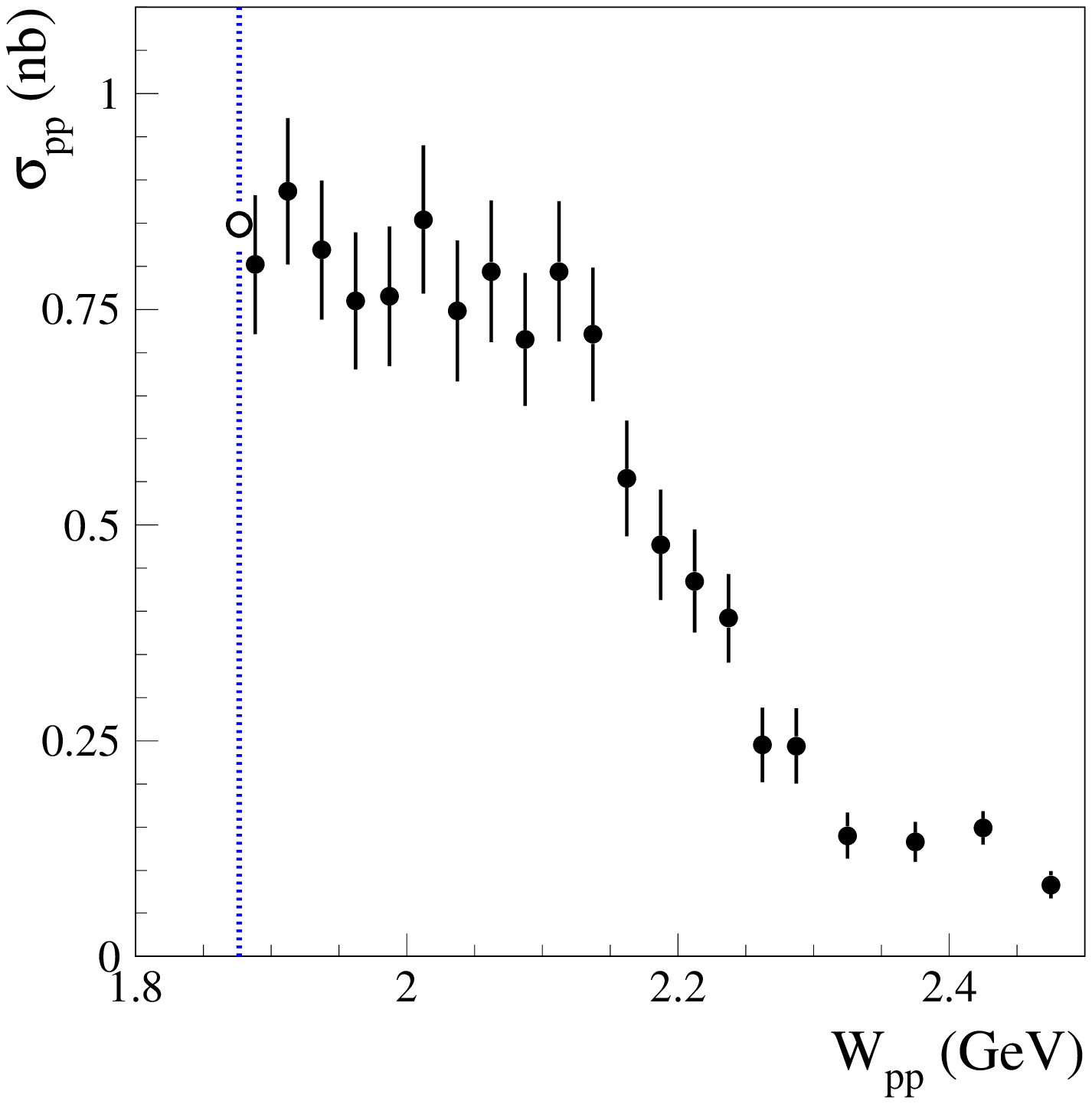,width=73mm}\vspace{-3mm}
\caption{\label{fig:pp} The \pp\ cross section obtained by \bbr\ (solid circles).
The dotted line represents the threshold and the empty circle is the Coulomb-enhanced 
threshold cross section, assuming $|G^p (4M_{p}^2)|= 1$.}
\ec\em
\ec
\vspace{-4mm}
\efi\\
Concerning P and D waves a further degree of approximation 
could be applied by means of the derivative at the origin or 
by means of a different approach~\cite{voloshin}.\\
The \eepp\ cross section~\cite{pp} in Fig.~\ref{fig:pp} shows the following 
peculiar features:
\begin{itemize}
\item it is suddenly 
different from zero at threshold, being $0.85 \pm  0.05$ \nb\ (by the way it is the 
only endothermic process  that has shown this peculiarity);
\item it is flat above threshold, within the experimental errors, in a c.m. energy 
interval of about 200 MeV  and then drops abruptly.
\end{itemize}
The expected Coulomb-corrected cross section at threshold is [see eq.~(\ref{eq:sigma})]
\bea
\sigma_{\pp}(4M_p^2) = \frac{\pi^2\alpha^3}{2M_p^2} \cdot |G^p(4M_{p}^2)|^2  
= 0.85 \cdot |G^p(4 M^2_p)|^2 \;\nb, \no
\eea
in striking similarity with the measured one, if $|G^p (4M_{p}^2)|= 1.00\pm 0.05$, 
and Coulomb interaction dominates the energy region at threshold.
\newline
Above threshold $|G_S^p|$ and $|G_D^p|$ have already  been achieved~\cite{noi3},  according to 
eq.~(\ref{eq:gsgd}) by means of the proton angular distribution and using a dispersion relation 
procedure applied to space-like and time-like data on the ratio $|G_E^p/G_M^p|$. Dispersion relations are 
needed to have access to the relative phase between  $\gs^p$ and $\gd^p$. In the c.m. energy 
range where the cross section is flat, real $\gs^p$ and $\gd^p$, as well as $\gd^p$ negative came out.
For the purposes of this letter this result is kept and  $\gs^p$ and $\gd^p$ are reevaluated more
exactly
according to eq.~(\ref{eq:sigma2}) [however the outcome is essentially the same achieved according to
 eq.~(\ref{eq:sigma})].
\bfi[h!]
\bc
\bm{85mm}\bc
\epsfig{file=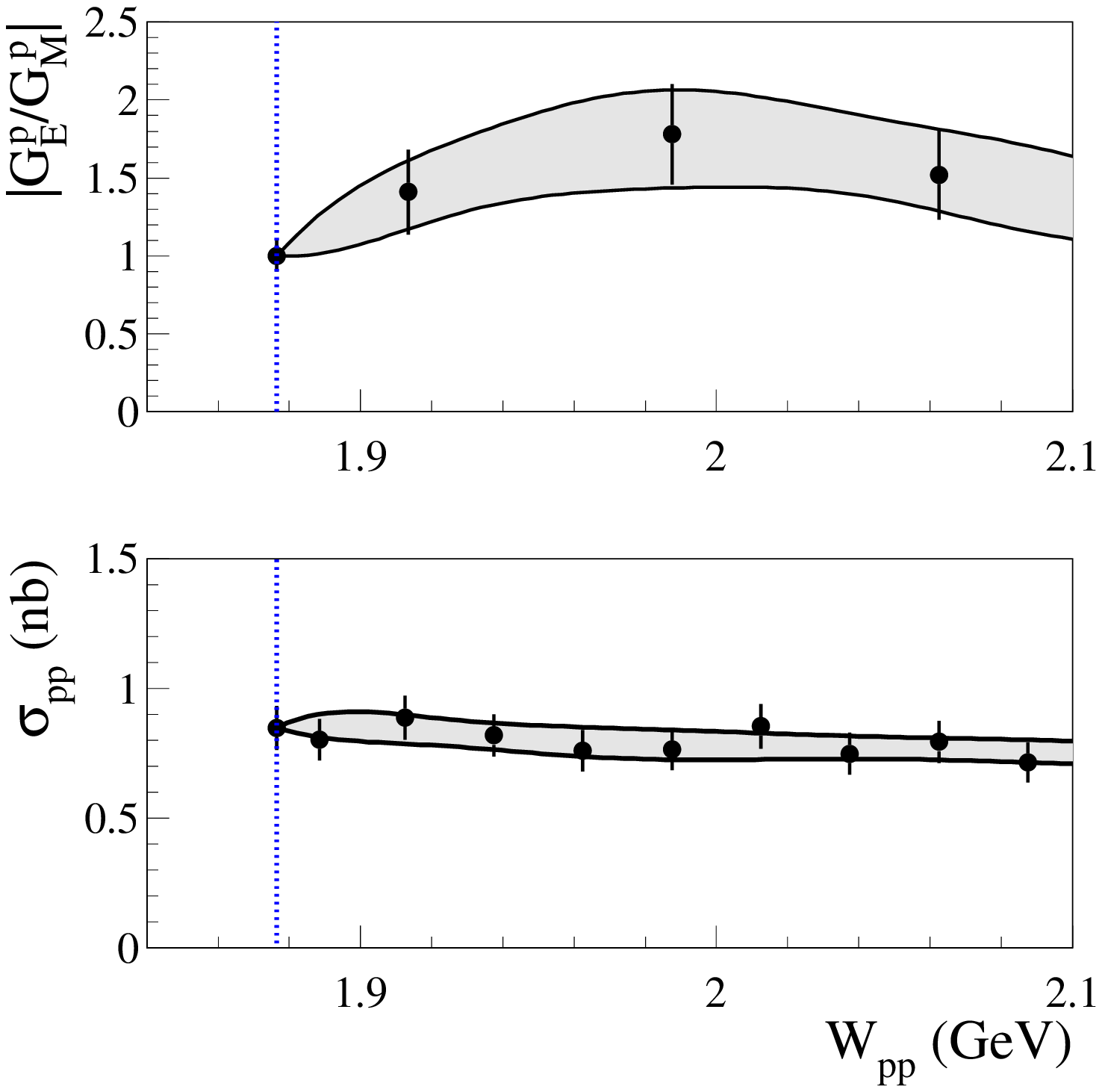,width=73mm}\vspace{-3mm}
\caption{\label{fig:rap} Top figure: modulus of the ratio $|\ge^p/\gm^p|$;
bottom figure: \pp\ cross section. The gray bands are the
fits and the dotted line is the threshold.}
\ec\em\ec
\efi\\
Figure~\ref{fig:rap} shows the near-threshold data on the 
ratio $|\ge^p/\gm^p|$ and \pp\ total cross section that have been 
used to obtain new $|G_S^p|$ and $|G_D^p|$, as shown in fig.~\ref{fig:gsgd},
where $|\gs^p|$ is compared to the inverse of the square root of the 
resummation factor of eq.~(\ref{eq:resum}) and it turns out 
\be
|\gs^p(4M_p^2\div 4\,\gev^2)|\simeq \frac{1}{\sqrt{\mathcal{R}}}=\sqrt{1-\exp(-\pi\alpha/\beta)}\,.
\no
\en
The agreement, within the errors, is striking. In other 
words: if the resummation factor is introduced in the Coulomb factor then the inverse of the
resummation factor is 
demanded in  $|G_S^p|^2$ by the data, strongly suggesting that it is an unnecessary factor.
If the resummation factor is not taken into account, it is $|G_S^p| \sim 1$ in a $\sim 200$ MeV 
 c.m. energy interval above threshold and then it drops abruptly.
\\
This conclusion could have been already foreseen on the basis of the flat cross section 
above threshold, unrelated to the expected steep increase above threshold due to the 
phase space.
There must be a cutoff  to the Coulomb dominance, but, what proton data are showing is that 
the energy scale for a baryon pair is hundred times greater than that expected for pointlike
charged fermions.
\bfi[h!]
\bc
\bm{120mm}\bc
\epsfig{file=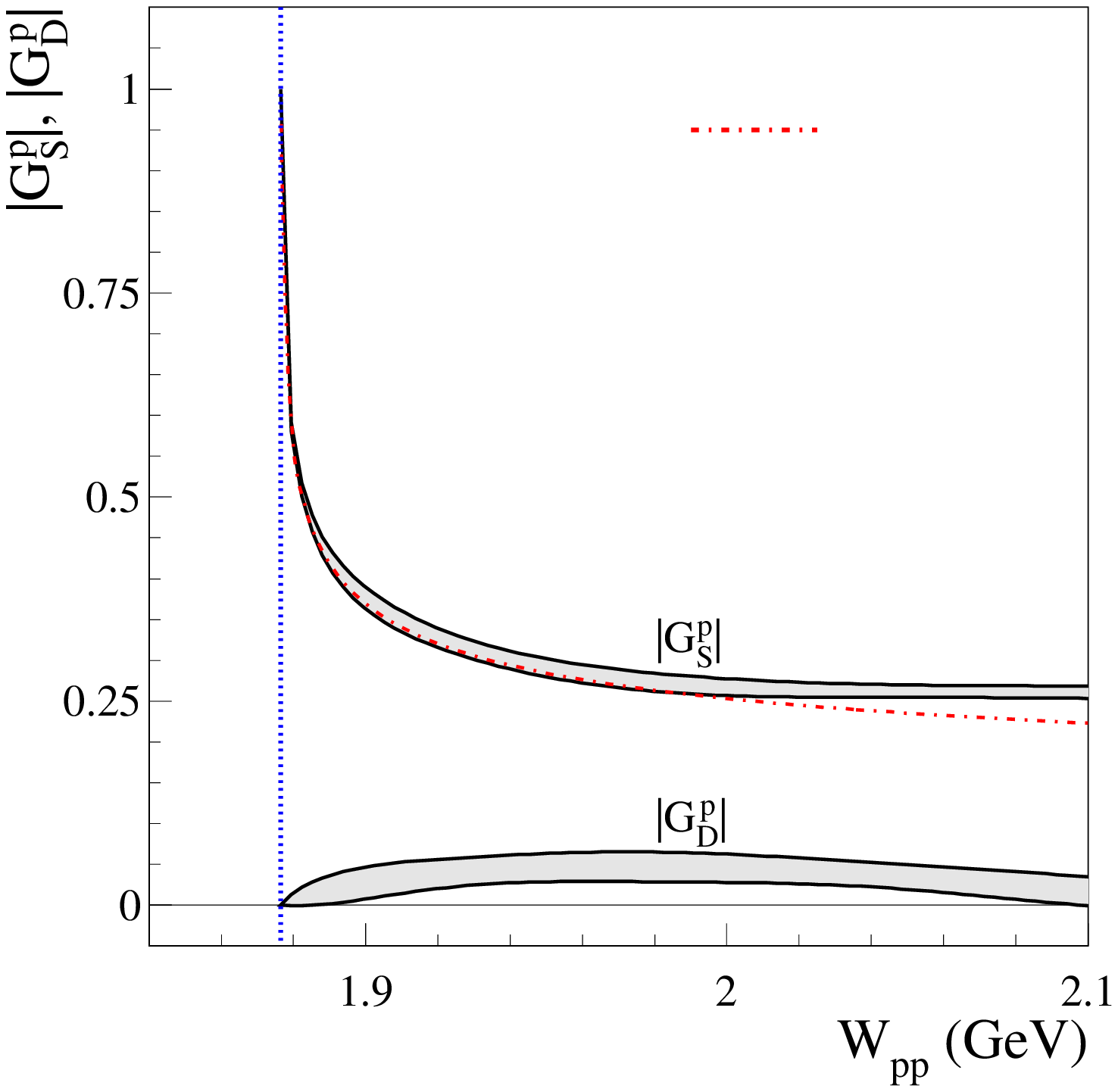,width=110mm}
\put(-78,265){\huge$\frac{1}{\sqrt{\mathcal{R}}}$}\vspace{-3mm}
\caption{\label{fig:gsgd} $|\gs^p|$ and $|\gd^p|$ obtained using
the ratio $|\ge/\gm|$, the total \pp\ cross section, and assuming a relative
phase $\phi=\pi$, see text. The dot-dashed curve is the inverse of the
square root of the resummation factor of eq.~(\ref{eq:resum}).}
\ec\em
\ec
\efi\\
Also in the case of \eelclc, see fig.~\ref{fig:lclc}, as already pointed out~\cite{noi2}, 
the cross section  measured by the Belle Collaboration~\cite{Belle} is not vanishing 
at threshold. If there is no 
resummation factor there is no major dependence on the mass resolution  and the 
expected cross section at threshold can be directly compared to the 
data, once the Coulomb enhancement is taken into account as well as assuming $|G^{\lc}(4 M^2_{\lc})| =1$.
There is a fair agreement, within the errors. 
\\
Measuring \eepp\ \bbr\ has also measured the cross section of \eenns\ + c.c., being  a 
significant background to \eepp.
To get from these data a cross section at threshold, a procedure has been exploited to 
avoid $N(1440)$ finite-width effects. The $N(1440)$ width as well as the \eenns\ + c.c.  are simulated. 
For each simulated event the $N(1440)$ momentum is evaluated and a new c.m. 
energy is achieved assuming a zero width.
The cross section obtained in this way is compared in fig.~\ref{fig:nstar}  to the pointlike cross section,
 Coulomb enhanced at threshold. Again there is agreement, suggesting that at threshold 
baryon pair production cross section behaves in a universal way.
\\
At last in the case of \eell, being $\Lambda$ a neutral baryon, final state Coulomb effects 
should not be taken into account and a finite cross section at threshold is not expected. 
Nevertheless the $\eell$ cross section data (Fig.~\ref{fig:ll}) show a 
threshold  behavior similar to that of $\sigma_{\pp}$.
\bfi[h!]
\bc
\bm{73mm}
\epsfig{file=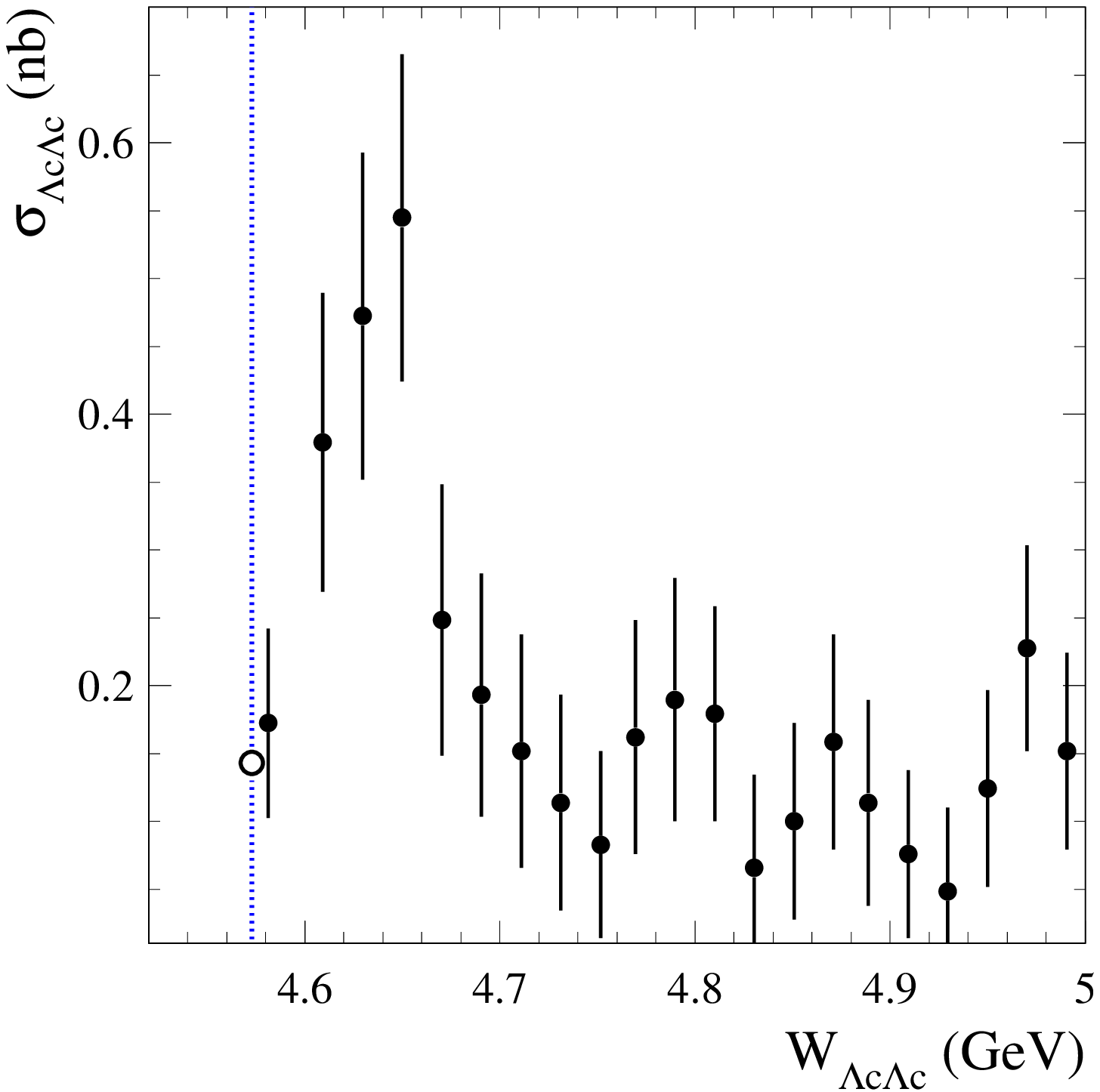,width=73mm}\vspace{-3mm}
\caption{\label{fig:lclc} The \lclc\ cross section obtained by Belle (solid circles).
The dotted line represents the threshold and the empty circle
is the cross section value Coulomb enhanced at threshold, assuming 
$|G^\lc (4M_{\lc}^2)|~=~1$.}
\em\hfill
\raisebox{-2.3mm}{\bm{73mm}
\epsfig{file=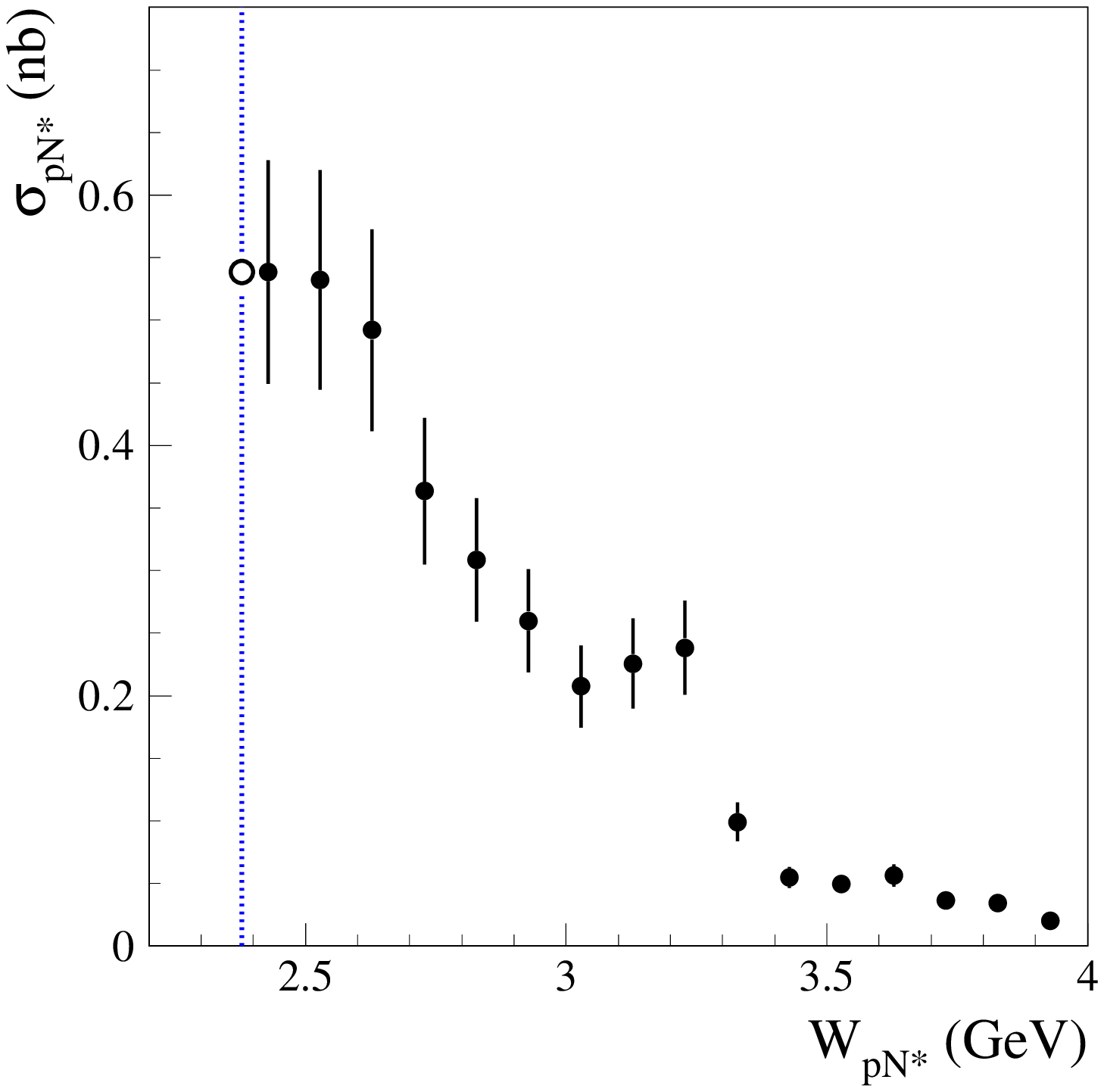,width=73mm}\vspace{-3mm}
\caption{\label{fig:nstar} The \eenns\ + c.c. cross section obtained by \bbr\ 
(solid circles). The dotted line represents the threshold and the empty circle
is the cross section value Coulomb enhanced at threshold, assuming 
$|G^{pN(1440)}[(M_p~\!~+~\!M_{N(1440)})^2]|= 1$.}
\em}
\ec
\efi\\
The cross sections $\sigma_{\sisi}$ and $\sigma_{\ls}$ have been measured by the 
\bbr\ Collaboration for the first time~\cite{ll}, although with very large errors as well
as \eennb\ measured by the FENICE Collaboration~\cite{FENICE} . The cross sections concerning
strange baryons are consistent with U-spin invariance expectation~\cite{park}, 
obtained under the assumption of negligible electromagnetic transitions between U-spin 
triplet and singlet~\cite{noi1}.
\\
Remnants of Coulomb interactions at the quark level have been assumed to explain non 
vanishing cross sections at threshold in the case of neutral baryon pair production.
\\
However, vanishing cross sections at threshold, rising according to the baryon velocity 
phase space factor, cannot be excluded according to the present \bbr\ data, as shown in 
fig.~\ref{fig:ll}. A much  better accuracy is needed to settle this 
issue.
\\
In conclusion, in the case of  \eepp\  near threshold as measured by \bbr\ it has been 
shown that, while at threshold there is the expected Coulomb enhancement factor in 
agreement with the non vanishing cross section, but $|G^p(4M_p^2)|=1$, above threshold 
there is no resummation factor. Other charged baryon pair cross sections \eebb\ show 
near threshold a similar behavior, within the errors.
\bfi[ht]
\bc
\bm{85mm}\bc
\epsfig{file=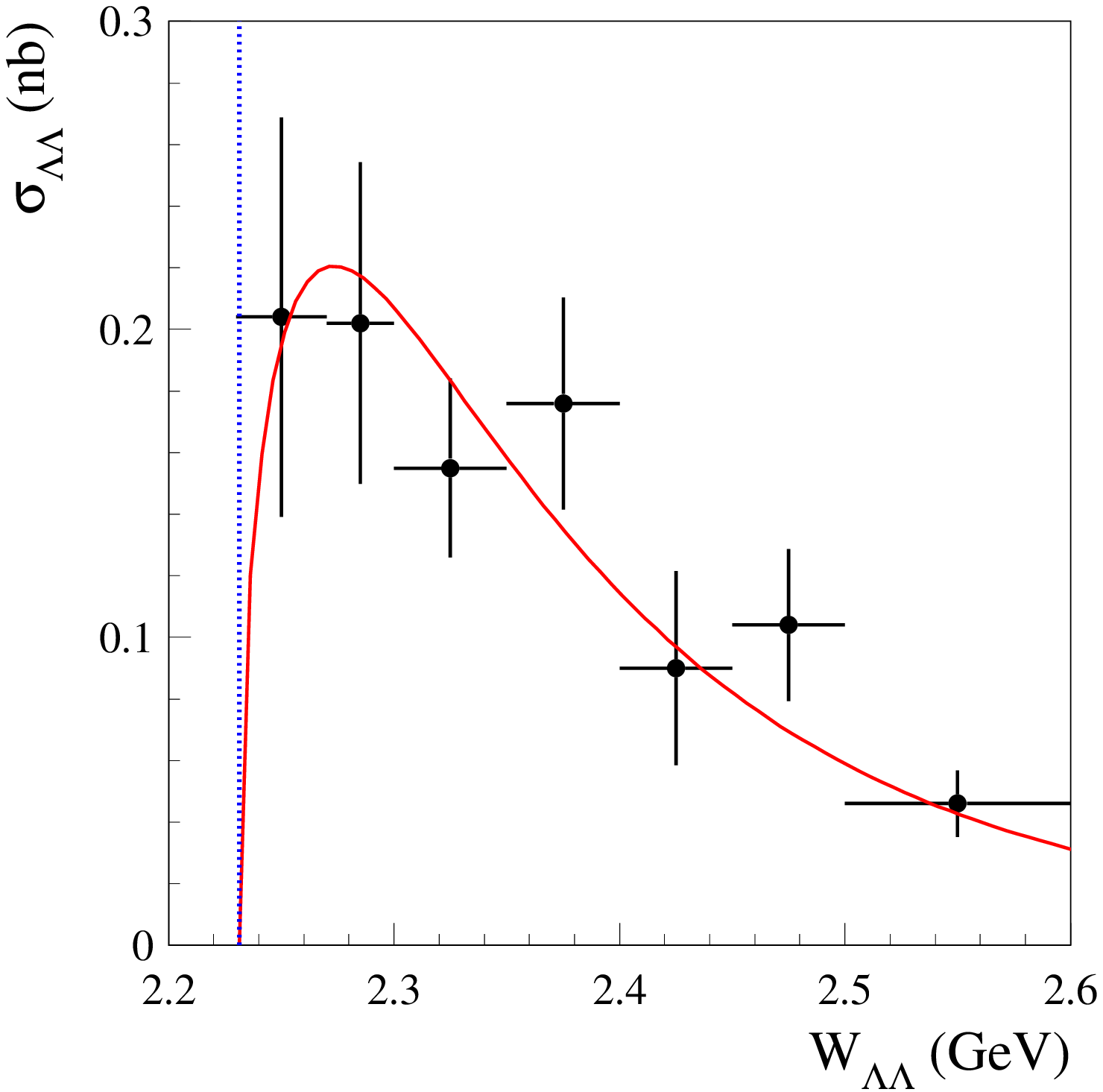,width=73mm}\vspace{-3mm}
\caption{\label{fig:ll} The \ll\ cross section obtained by \bbr.
The dotted line represents the threshold and 
a fit is reported, obtained including the
outgoing baryon velocity factor and by means of an effective FF behaving like
a dipole.} 
\ec\em
\ec
\efi\\
New data near threshold are coming from CMD2 and SND at VEPP2000 and on a larger interval
from BESIII at BEPCII by means of initial state radiation, as well as a test of the Sommerfeld
rescattering formula on a pure QED process, like \eett. 
\\
The investigation of the time-like behavior of nucleon FF's has been carried out
by many authors using different approaches, models and phenomenological descriptions;
in Ref.~\cite{timelike} we report only an incomplete list. However, the
result we present in this letter is a pure statement of fact and hence completely
model-independent. Possible interpretations and phenomenological explanations
are under study~\cite{noi4}.
\acknowledgments
We warmly acknowledge Antonino Zichichi, Giulia Pancheri and Yogi Srivastava for 
their strong support and the 
\bbr\ Collaboration, in particular Budker Institute colleagues, who achieved most of the 
experimental results. 
%
%
%
%

\end{document}